\documentclass[apl,aip,groupedaddress,twocolumn,reprint,showpacs,preprintnumbers,amsmath,amssymb]{revtex4-1}

\usepackage{graphicx}% Include figure files
\usepackage{bm}% bold math
\usepackage{epstopdf}

\begin{document}
%\preprint{APS/123-QED}

\title{Indirect Coupling between Two Cavity Photon Systems via Ferromagnetic Resonance}
\author{Paul Hyde}
\email{umhydep@myumanitoba.ca}
\author{Lihui Bai}
\email{bai@physics.umanitoba.ca}
\author{Michael Harder}
\author{Christophe Match}
\author{Can-Ming Hu}

\affiliation{Department of Physics and Astronomy, University of Manitoba, Winnipeg, Canada R3T 2N2}

\date{\today}

\begin{abstract}

We experimentally realize indirect coupling between two cavity modes via strong coupling with the ferromagnetic resonance in Yttrium Iron Garnet (YIG). We find that some indirectly coupled modes of our system can have a higher microwave transmission than the individual uncoupled modes. Using a coupled harmonic oscillator model, the influence of the oscillation phase difference between the two cavity modes on the nature of the indirect coupling is revealed. These indirectly coupled microwave modes can be controlled using an external magnetic field or by tuning the cavity height. This work has potential for use in controllable optical devices and information processing technologies.

\end{abstract}

\maketitle

%\section{Introduction}

The indirect coupling of cavity modes via a waveguide has been studied theoretically and experimentally for use in optical information processing\cite{1}. This indirect coupling dramatically modifies the transmission spectra, and is widely used for optical filtering, buffering, switching, and sensing in photonic crystal structures\cite{2,3,4,5}. For micro/nano disk optical cavities, coupling properties are determined by the spatial distance between the disk and the waveguide during the fabrication process. Therefore, a tunable coupling between indirectly coupled cavity modes is required for potential applications.

\begin{figure}[!]
\includegraphics[width=\linewidth]{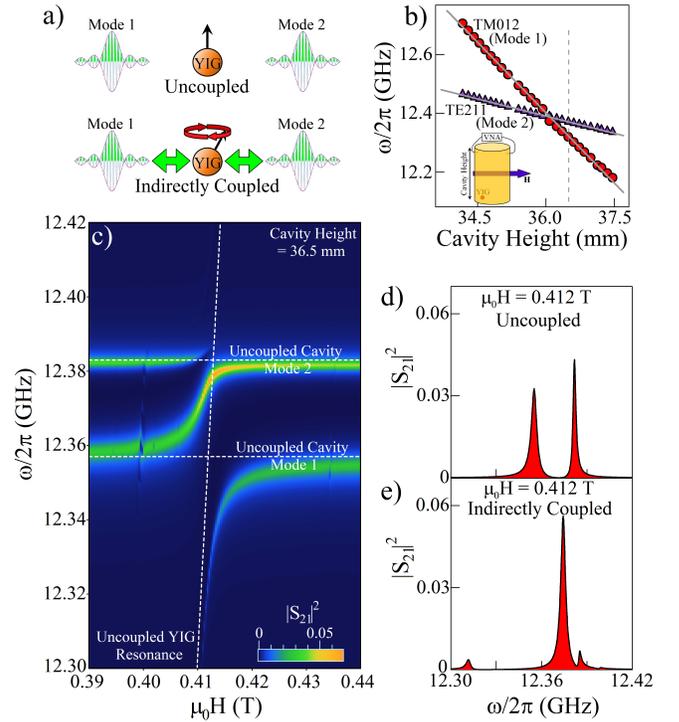}
\caption{(Colour online) \textbf{(a)} In an uncoupled system individual elements do not interact with each other. In our experimental system a YIG sphere simultaneously couples to two separate cavity modes, indirectly coupling the modes together. \textbf{(b)} The frequencies of the two cavity modes are functions of the height of the cylindrical microwave cavity and cross near a height of 36 mm. The inset shows a sketch of the microwave cavity.  \textbf{(c)} Transmission spectrum $S_{21}$ of our indirectly coupled system, as a function of the external magnetic field at a microwave cavity height of 36.5 mm [dashed line in (b)]. \textbf{(d)} Transmission spectrum $S_{21}$ of our cavity system, with a height of 36.5 mm at an external field $\mu_{0}H=0.412$ T, in an uncoupled state (NO YIG in cavity) and \textbf{(e)} an indirectly coupled state (YIG in cavity), showing the influence of coupling on the resonant modes.}
\label{Fig 1}
\end{figure}

Recently, strong coupling between a microwave cavity mode and ferromagnetic resonance (FMR) has been realized at room temperature\cite{6,7,8,9,10,11,12,13,14,15,16,17}. Exchange interactions lock the high density of spins in YIG into a macro-spin state, leading to strong coupling with a cavity mode which can be adjusted using an external magnetic field. Potential applications of this form of strong coupling are currently being explored. For example, indirect coupling between the FMR in two YIG spheres has produced dark magnon modes with potential uses in information storage technologies\cite{18}, and the FMR of YIG has been indirectly coupled with a qubit through a microwave cavity mode\cite{19}. Instead of using a microwave cavity mode to build a bridge between two oscillators, we have used the FMR in YIG to produce indirect coupling between two cavity modes.

In this work, we present two cavity modes which indirectly couple via their strong coupling with the FMR in YIG at room temperature. The two cavity modes are labelled $h_{\omega 1}(\omega)$ and $h_{\omega 2}(\omega)$ respectively, and are independent of each other when there is no direct coupling between them. Here $\omega_{1}$ and $\omega_{2}$ are the uncoupled resonance frequencies of each cavity mode and $\omega$ is the input microwave frequency. The two cavity modes can be indirectly coupled with each other when they both individually interact with the FMR in YIG and this indirect coupling can be controlled using an external magnetic field. We found that the microwave transmission properties change dramatically for the coupled modes as the external field is tuned. Our experimental results, together with an extended coupled harmonic oscillator model, demonstrate the nature of indirect coupling and coherent information transfer. This tunable interaction between orthogonal cavity modes could potentially be used to build controllable optical and microwave devices.

The microwave cavity used in our experiment was made of oxygen-free copper with a height tunable cylindrical structure. The diameter of the cavity is 25 mm and the height is tunable in a range between 24 mm and 45 mm. Although multiple modes can exist inside of the cavity, the TM$_{012}$ mode (with a cavity frequency of $\omega_{1}$) and the TE$_{211}$ mode (with a cavity frequency of $\omega_{2}$) were chosen to demonstrate indirect coupling in this work. With no YIG inside of the cavity, the microwave transmission, $S_{21}$, was measured using a Vector Network Analyser (VNA) as a function of frequency. The output microwave power of the VNA is 1 mW. The amplitude of the transmission is proportional to the resonance amplitudes of both cavity modes at a given microwave frequency, $|S_{21}(\omega)|^{2}\propto|h_{\omega 1}(\omega)+h_{\omega 2}(\omega)|^{2}$. Here, $h_{\omega 1}=\frac{\Gamma_{1}\omega^{2}}{\omega^{2}-\omega_{1}^{2}+2i\beta_{1}\omega_{1}\omega}h_{0}$ and $h_{\omega 2}=\frac{\Gamma_{2}\omega^{2}}{\omega^{2}-\omega_{2}^{2}+2i\beta_{2}\omega_{2}\omega}h_{0}$ are the response functions of each cavity mode near the resonance conditions. $\omega_{1}$, $\omega_{2}$, $\beta_{1}$, and $\beta_{2}$ are the cavity mode resonance frequencies and damping. $\Gamma_{1}$ and $\Gamma_{2}$ denote the impedance matching parameters for each cavity mode. $h_{0}(\omega)$ is the microwave field used to drive resonance in the cavity and is eliminated by normalization in the microwave transmission. The microwave transmission spectra with no YIG in the cavity allows the individual cavity mode frequencies and damping to be evaluated. Fig. 1(b) plots the resonant frequencies of $\omega_{1}$ and $\omega_{2}$ as a function of the height of the microwave cavity, both agree well with the solutions for Maxwell's equations (solid lines) in a cylindrical microwave cavity. That the two cavity modes cross each other indicates that there is no direct coupling between them. The different microwave magnetic field distributions of the two modes inside the cavity leads to them having different coupling strengths with the FMR in YIG. For a given cavity height of 36.5 mm, the parameters of the two cavity modes were determined to be:  $\omega_{1}/2\pi=12.357$ GHz, $\beta_{1}=1.9\times 10^{-4}$, $\Gamma_{1}=6.1\times 10^{-5}$, $\omega_{2}/2\pi=12.382$ GHz, $\beta_{2}=0.91\times 10^{-4}$, and $\Gamma_{2}=3.7\times 10^{-5}$.

A YIG sphere\cite{20} placed inside the cavity allows for indirect coupling between the two cavity modes. The YIG sphere has a diameter of 1 mm, saturation magnetization $\mu M_{0}=0.178$ T, gyromagnetic ratio $\gamma=28\times 2\pi\mu_{0}$ GHz/T, and Gilbert damping $\alpha=1.15\times 10^{-4}$. The YIG sphere was placed at the bottom of the cavity near the wall as shown in the inset of Fig. 1(b). An external magnetic field, \textbf{H}, was applied to the YIG as shown in the inset. This magnetic field allows us to tune the FMR frequency of the YIG, $\omega_{FMR}$, following an $\omega$-H dispersion $\omega_{FMR}=\gamma(H+H_{Ani})$. Here, the anisotropy field of the sphere is $\mu_{0}H_{Ani}=0.0294$ T.

Transmission measurements of our coupled system are plotted in Fig. 1(c), which shows the amplitude $|S_{21}|^{2}$ as a function of the input microwave frequency ($\omega$) and the external magnetic field \textbf{H}. By increasing the \textbf{H} field, the FMR frequency ($\omega_{FMR}$) first increases to the lower cavity mode frequency $\omega_{1}$, then reaches the higher cavity mode frequency $\omega_{2}$ as indicated by the dashed lines. By doing this, the two cavity modes are indirectly coupled together via their direct coupling with the FMR in YIG, producing three coupled modes. We observed a maximum in the microwave transmission amplitude when the middle mode (later labelled Mode B) crosses the dispersion of the YIG FMR (dashed line) due to the resonances of the two cavity modes being in-phase. Fig. 1(d) and (e) show how the addition of the YIG sphere into the cavity affects the observed resonant modes at an external field of 0.412 T; with both the number and position of the observed modes changing once the sphere is placed in the cavity.

\begin{figure}[t]
\centering
\includegraphics[width=\linewidth]{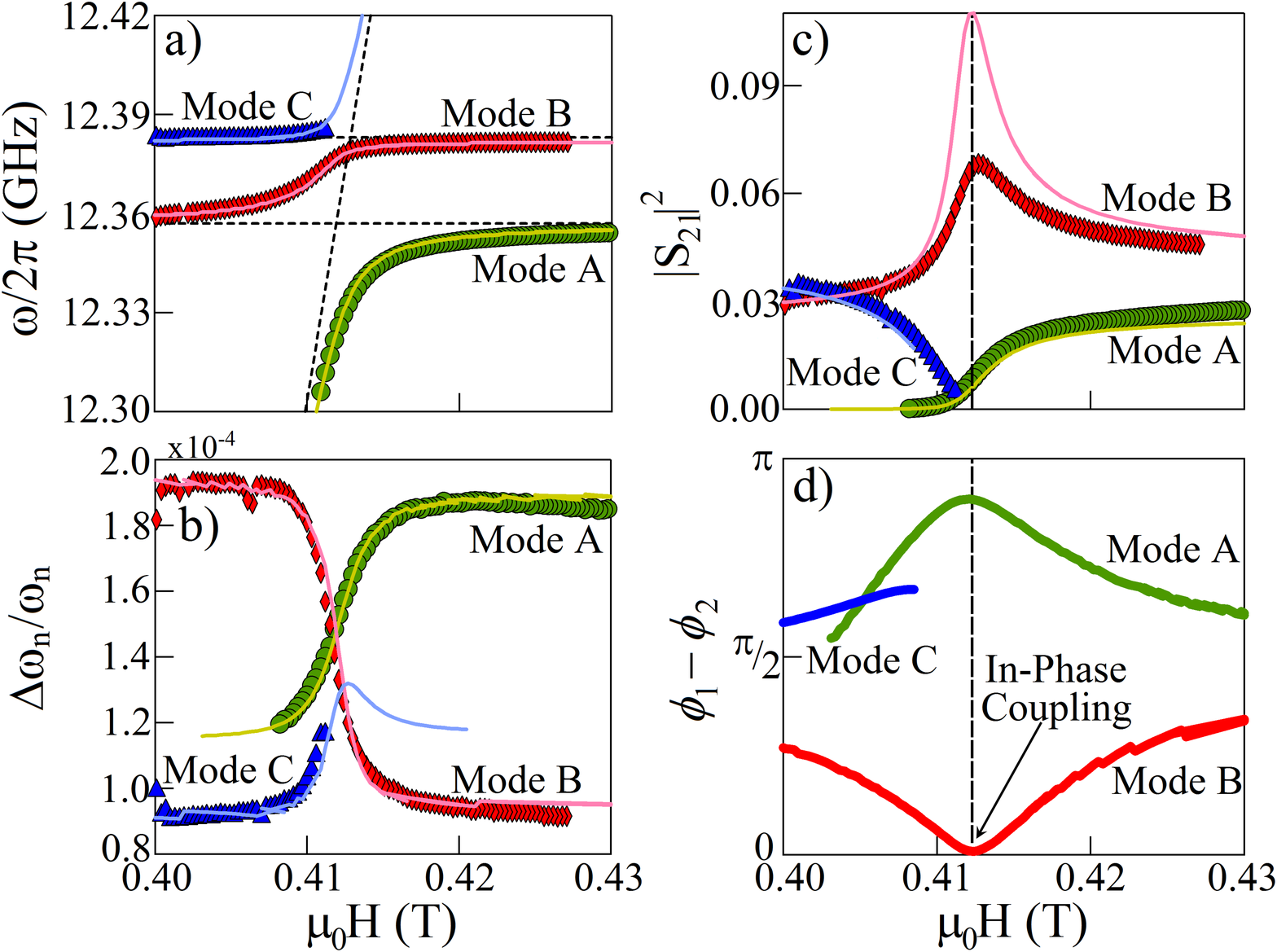}
\caption{(Colour online) \textbf{(a)} and \textbf{(b)} display the $\omega$-H dispersion and damping evolution (symbols) of each of the Normal Modes in our system. They are compared to calculations from Eq. 1 (solid curves). \textbf{(c)} The amplitudes of the Normal Modes, $|S_{21}|^{2}$, are dramatically enhanced or suppressed during coupling. \textbf{(d)} The relative phase between the two cavity modes, $\phi_{1}-\phi_{2}$, was calculated during indirect coupling. The in-phase point of Mode B corresponds to its maximum amplitude in (c).}
\label{Fig 2}
\end{figure}

To further understand the nature of this indirect coupling between the two cavity modes, an expanded coupled harmonic oscillator system is used to calculate coupling features including the $\omega$-H dispersion, damping evolution, and amplitudes. Coupled harmonic oscillators have previously been used to accurately model strong coupling between a cavity mode and FMR in YIG\cite{21}. The coupling strengths between each cavity mode and the FMR in YIG, $\kappa_{1}=0.070$ and $\kappa_{2}=0.043$, were evaluated using the two coupled harmonic oscillator model when the two cavity mode frequencies were well separated (not shown here). The local microwave magnetic field distribution of each mode, with respect to the external field orientation, lead to different coupling strengths between each cavity mode and the FMR in YIG\cite{22}. A slight change of the cavity height does not change the coupling strength of each mode. However, the coupled system observed in this work can no longer be modelled by the two coupled harmonic oscillator model. To take into account the second cavity mode, a three oscillator system is considered rather than the two in Ref.$[21]$. Two of the oscillators describe the cavity modes with amplitudes of $h_{\omega 1}$ and $h_{\omega 2}$, each separately coupled with the third representing the FMR in YIG with amplitude $m$. Therefore, the indirect coupling model can be written in the form;

\begin{widetext}
\begin{equation}
\left( \begin{array}{ccc}
\omega^{2}-\omega_{1}^{2}+i2\beta_{1}\omega_{1}\omega & 0 & -\kappa_{1}^{2}\omega_{1}^{2} \\
0 & \omega^{2}-\omega_{2}^{2}+i2\beta_{2}\omega_{2}\omega & \kappa_{2}^{2}\omega_{2}^{2} \\
-\kappa_{1}^{2}\omega_{1}^{2} & \kappa_{2}^{2}\omega_{2}^{2} & \omega^{2}-\omega_{FMR}^{2}+i2\alpha\omega_{FMR}\omega \end{array} \right) \left( \begin{array}{c}
h_{\omega 1} \\
h_{\omega 2} \\
m \end{array} \right) = \omega^{2} \left( \begin{array}{c}
\Gamma_{1} \\
\Gamma_{2} \\
0 \end{array} \right) h_{0}
\end{equation}
\end{widetext}

\begin{figure}[b]
\centering
\includegraphics[width=\linewidth]{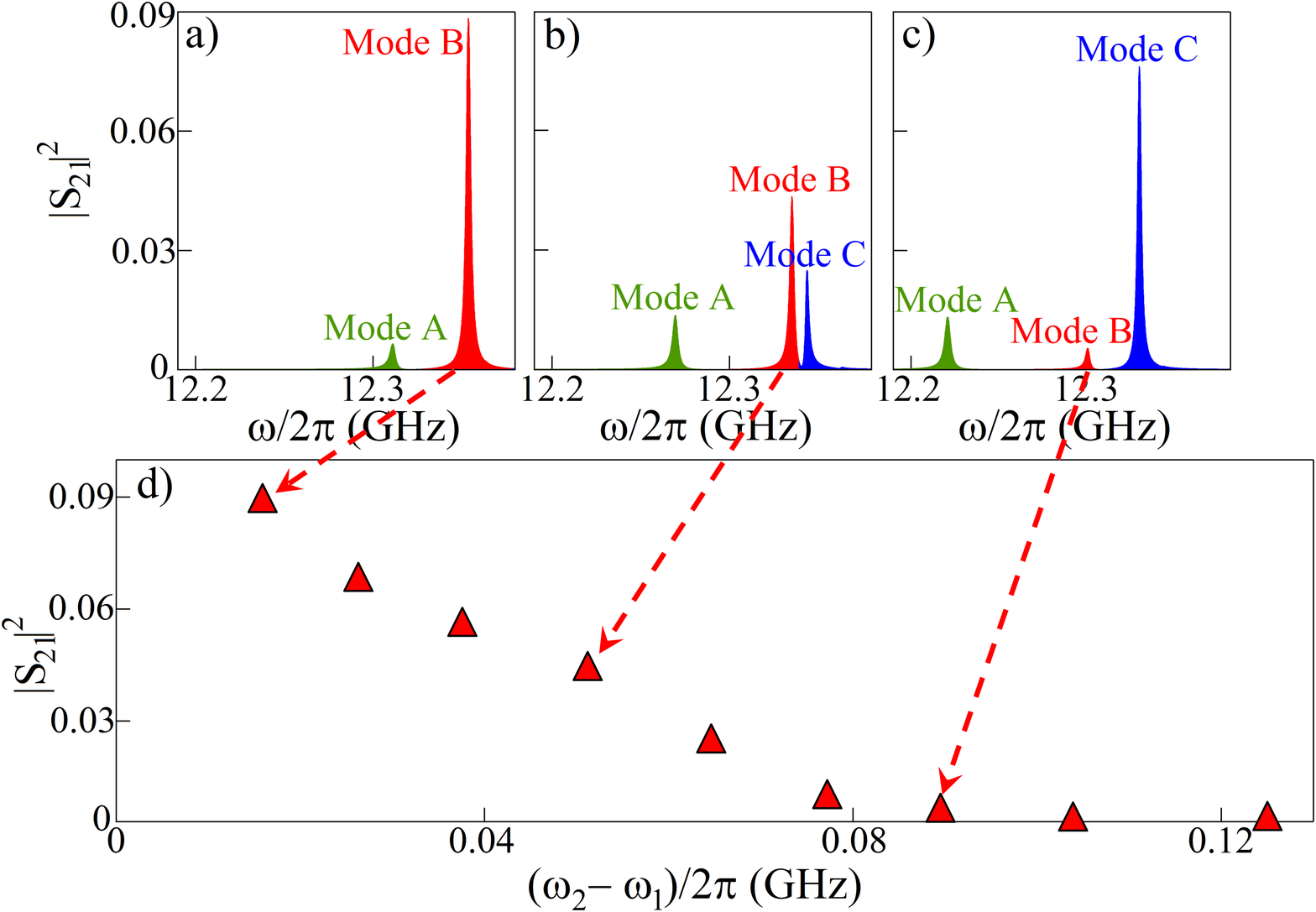}
\caption{(Colour online) \textbf{(a)}, \textbf{(b)}, and \textbf{(c)} show the transmission spectrum of the three Normal Modes for different $\omega_{2}-\omega_{1}$ values. \textbf{(d)} Amplitude of the in-phase point of Mode B as a function of $\omega_{2}-\omega_{1}$.}
\label{Fig 3}
\end{figure}

Here the diagonal terms are the uncoupled resonance conditions of the two cavity modes and the FMR in YIG. The off-diagonal terms are the coupling strengths. The two zeros indicate that there is no direct coupling between the two cavity modes. To explain our experimental observations we must include a $\pi$-phase delay between the resonance frequencies of neighbouring cavity modes, although the physical source of this phase shift is still an open question. This is the source of the additional minus sign in the $\kappa_{1}$ terms. Eq. 1 allows us to predict the characteristics of indirect coupling between cavity modes via FMR in YIG.

By finding the complex eigen-frequencies $\omega_{n}$ (n = A, B, C, denoting the Modes labelled in Figure 2) of the coupling matrix at a given \textbf{H} field, we can plot the calculated resonance frequency $Re(\omega_{n})$, and the normalized line width $|Im(\omega_{n})|/Re(\omega_{n})$ in Fig. 2(a) and (b) using solid curves. This matches the observed $\omega$-H dispersion and damping evolution seen in the measurements (symbols).

Furthermore, we are able to calculate the amplitude and relative phase of the microwave transmission $h_{\omega 1}$, $h_{\omega 2}$, and $m$ using Eq. 1. The calculated transmission amplitude $|S_{21}|^{2}$ was plotted in Fig. 2(c) (solid curves) and compared with that from our experimental results (symbols). An amplitude peak is seen in both the experimental results and the theoretical calculation. The phase difference ($\phi_{1}-\phi_{2}$) between $h_{\omega 1}$ and $h_{\omega 2}$ is calculated and plotted in Fig. 2(d). The applied field strength at the maximum amplitude corresponds to an in-phase point, highlighted in Fig. 2(d), where the phases of the two cavity modes $h_{\omega 1}$ and $h_{\omega 2}$ are equal ($\phi_{1}-\phi_{2}=0$). Meanwhile the amplitude decrease of the other Normal Modes is due to the relative phase difference between the two cavity modes approaching $\pi$. Hence, coherent phase control between two indirectly coupled cavity modes is detected through amplitude enhancement of the microwave transmission and explained by our three oscillator model.

The in-phase point observed occurs when Mode B crosses the uncoupled dispersion of the YIG FMR. The amplitude of this in-phase point also depends on the difference between the resonant frequencies of the two cavity modes ($\omega_{2}-\omega_{1}$). By tuning the cavity height, the in-phase point can be measured for different values of $\omega_{2}-\omega_{1}$ as shown in Fig. 3(a), (b), and (c). The amplitude of these in-phase points, highlighted in red, increases when the two cavity mode frequencies are near to each other. As summarized in Fig. 3(d), the transmission amplitude $|S_{21}|^{2}$ decreases as the two cavity mode frequencies are separated. Therefore, the microwave transmission of the in-phase point can also be controlled by the cavity height.

In summary, we experimentally demonstrate controllable indirect coupling between two microwave cavity modes through a YIG sphere. The coupling features are analysed and explained using a three coupled harmonic oscillator model. Microwave modes produced due to indirect coupling were observed to have a higher transmission rate than the two uncoupled cavity modes. We also demonstrated that these indirectly coupled modes can be controlled with an external field and by changing the cavity's height. Therefore, due to the controllable nature of our findings, our work can be useful for designing new optical devices for information processing.

The authors would like to thank B. Yao for useful discussions. P.H. is supported by the UMGF program. M.H. is supported by an NSERC CGSD Scholarship. This work has been funded by NSERC, CFI, and NSFC (No. 11429401) grants (C.-M. Hu).

\end{document}